**Title**: *An automated and time-efficient framework for simulation of coronary blood flow under steady and pulsatile conditions*

**Authors**: Guido Nannini[1], Simone Saitta[1], Luca Mariani[1], Riccardo Maragna[2], Andrea Baggiano[2,3], Saima Mushtaq[2], Gianluca Pontone[2,4], Alberto Redaelli[1]

**Affiliations**:

1. Department of Electronics Information and Bioengineering, Politecnico di Milano, Milan, Italy
2. Department of Perioperative Cardiology and Cardiovascular Imaging D, Centro Cardiologico Monzino IRCCS, Milan, Italy
3. Department of Clinical Sciences and Community Health. University of Milan, Milan, Italy
4. Department of Biomedical, Surgical and Dental Sciences, University of Milan, Milan, Italy

**Abstract**. *Background and Objective*: Invasive fractional flow reserve (FFR) measurement is the gold standard method for coronary artery disease (CAD) diagnosis. FFR-CT exploits computational fluid dynamics (CFD) for non-invasive evaluation of FFR, simulating coronary flow in virtual geometries reconstructed from computed tomography (CT), but suffers from cost-intensive computing process and uncertainties in the definition of patient specific boundary conditions (BCs). In this work, we investigated the use of time-averaged steady BCs, compared to pulsatile to reduce the computational time and deployed a self-adjusting method for the tuning of BCs to patient-specific clinical data. *Methods*: 133 coronary arteries were reconstructed form CT images of patients suffering from CAD. For each vessel, invasive FFR was measured. After segmentation, the geometries were prepared for CFD simulation by clipping the outlets and discretizing into tetrahedral mesh. Steady BCs were defined in two steps: *i)* rest BCs were extrapolated from clinical and image-derived data; *ii)* hyperemic BCs were computed from resting conditions. Flow rate was iteratively adjusted during the simulation, until patient's aortic pressure was matched. Pulsatile BCs were defined exploiting the convergence values of steady BCs. After CFD simulation, lesion-specific hemodynamic indexes were computed and compared between group of patients for which surgery was indicated and not. The whole pipeline was implemented as a straightforward process, in which each single step is performed automatically. *Results*: Steady and pulsatile FFR-CT yielded a strong correlation ($r=0.988$, $p<0.001$) and correlated with invasive FFR ($r=0.797$, $p<0.001$). The per-point difference between the pressure and FFR-CT field predicted by the two methods was below 1% and 2%, respectively. Both approaches exhibited a good diagnostic performance: accuracy was 0.860 and 0.864, the AUC was 0.923 and 0.912, for steady and pulsatile case, respectively. The computational time required by steady BCs CFD was approximatively 30-folds lower than pulsatile case. *Conclusions*: This work shows the feasibility of using steady BCs CFD for computing the FFR-CT in coronary arteries, as well as its computational and diagnostic performance within a fully automated pipeline.

**List of acronyms.**

| | |
|---|---|
| 5WK | 5 Elements Winkessel Model |
| AUC | Area under the curve |
| BC | Boundary condition |
| CAD | Coronary artery disease |
| CCTA | Coronary computed tomography angiography |
| CFD | Computational Fluid Dynamics |
| DP | Diastolic pressure |
| FFR | Fractional flow reserve |
| FFR-CT | FFR measured with CFD |
| FFR-CT$_{SS}$ | Subscript 'SS' indicates Steady Simulation |
| FFR-CT$_{TR}$ | Subscript 'TR' indicates Transient Simulation |
| FFR-grad | FFR gradient per millimeter |
| HR | Heart rate |
| LAD | Left Anterior Descending Coronary Artery |
| LCX | Left Circumflex Coronary Artery |
| LoA | Limit of Agreement |
| LPM | Lumped parameter model |
| NSE | Navier-Stokes equations |
| Pa | Proximal pressure |
| Pao | Mean aortic pressure |
| PCI | Percutaneous coronary intervention |
| Pd | Distal pressure |
| PPG | Pullback Pressure Gradient |
| RCA | Right Coronary Artery |
| ROC | receiver operating characteristic |
| SP | Systolic pressure |
| TCRI | Total coronary resistance Index |
| VMTK | Vascular Modeling Toolkit |
| WSS | Wall Shear Stress |

## 1) Introduction

Myocardial ischemia is a severe cardiovascular disease, caused by the obstruction of coronary arteries, which leads to a restriction in blood supply in cardiac tissues. In Western Country, coronary artery disease (CAD) is the leading cause of death, causing approximately 650k deaths per year[1]. In the clinical practice, coronary computed tomography angiography (CCTA) has emerged as the foremost imaging modality for diagnosing CAD, offering high-resolution volumetric images of the entire heart, facilitating comprehensive analysis of coronary vessels. CCTA allows to identify adverse morphological characteristics associated with CAD progression, such as stenosis grading, plaque burden and vessel tortuosity[2], for effective risk stratification and preventive intervention. In particular, stenosis grading has been considered for many years as the main predictor of myocardial infarction[3,4]. In the last years, fractional flow reserve (FFR) has emerged as the gold standard for invasive functional assessment of CAD[5]. FFR is an indicator of the functional severity of a stenosis, which indirectly quantifies the ratio of flow rate across the stenosis with respect to the flow in the vessel in absence of the lesion. In practice, FFR is defined as the average ratio, over a full cardiac cycle, between distal pressure (Pd) and proximal pressure (Pa), measured during maximal hyperemia[6]. Percutaneous coronary intervention (PCI) is indicated for lesions with FFR<0.80[5]. Different trials proved not only the higher accuracy of FFR as a diagnosis index compared to the stenosis grading, achieving a 1/3 lower rate of major adverse events, but also the higher sensitivity in identifying obstruction associated with ischemia risk[7,8]. Despite such encouraging achievements, the uptake of FFR in clinical practice is still low worldwide, most likely due to the high cost associated to the medical equipment, the need of drug administration to induce maximal hyperemia, the required time to perform FFR measurement and the possible discomfort for the patients that may not endure such an invasive procedure.

Computational fluid dynamics (CFD) is a numerical method that allows *in silico* simulation of coronary flow, thus providing a non-invasive estimate of the FFR. By integrating anatomical information from CCTA and functional clinical data, it is possible to model blood flow and accurately compute the FFR-CT[9]. Compared to invasive FFR, the FFR-CT based CAD assessment entails lower costs, is non-invasive and does not require additional administration of vasodilators to induce hyperemia, ultimately being an enticing option for CAD diagnosis. An additional advantage of FFR-CT is the fact that, with a single simulation, it is possible to obtain flow and pressure maps in the whole coronary tree, thus identifying severe stenosis on different branches without requiring for multiple invasive acquisition. The reliability of FFR-CT as a strong independent predictor of ischemia was proved in different randomized trials[10,11].

The deployment of a reliable CFD model for computing FFR-CT, that realistically reproduces *in vivo* coronary hemodynamics, essentially requires three steps: *i)* the definition of an accurate 3D anatomical model; *ii)* the setting of patient-tailored boundary conditions (BCs); *iii)* the numerical solution of the Navier-Stokes equations (NSE). Each of these steps involves time-consuming operations that can be sped up and automated.

High-resolution CCTA images enable precise geometry reconstruction of coronary arteries, albeit the manual segmentation, conducted by an expert operator, is a time-consuming operation requiring considerable time. Deep learning algorithms have proved to be an effective tool for automating segmentation of medical images and once trained, inference generally requires few minutes[12]. Few commercial softwares exploit semi-automatic methods[13,14], nevertheless various fully automatic deep-learning based models for coronary artery segmentation have been detailed in the literature[15–19].

Defining patient-tailored BCs can be challenging as it generally relies on few available clinical data (e.g., brachial pressure, heart rate). Taylor et al.[20] were the among the firsts to achieve a CFD model for measuring FFR-CT. As BCs, they used a heart lumped parameter model (LPM) coupled at the inlet, and LPMs of distal impedances at each outlet. The parameters of each LPM used were tuned to patient-specific data, by solving the system of partial differential equations associated to a full-circulation closed loop, achieving to reproduce realistic flow and pressure waveform in coronary arteries. Subsequent studies have elaborated on comparable models to simulate coronary flow within a 3D domain[21–28], assessing the impact of using different BCs[29–31] and different vessel morphologies[26,32–34]. Nevertheless, patient-specific measures of coronary flow rate or aortic pressure, despite being feasible, are rarely available, and this implies making assumptions on inlet BCs, while outlet BCs can only be extrapolated, thus inevitably introducing an uncertainty in the model. Moreover, FFR is defined at maximal hyperemia, during which hemodynamics is altered due to the administration of vasodilators, to which patient-specific response in terms of changes in coronary flow and distal impedance is unknown. As a result, a proper tuning of BCs can take up a significant amount of time in the implementation of a coronary CFD model.

Finally, the computational cost required to run a full 3D pulsatile simulation of coronary flow makes such models extremely time demanding (up to 48 hours). The time constraints in the deployment of a coronary CFD model discourage its application in the clinical practice: therefore, increasing both efficiency and automation of *in silico* modeling and reducing the time-demand of each step in the model implementation, is crucial to achieve on-site FFR-CT assessment within a short time. Solutions to reduce the cost of the numerical simulation have been investigated mostly exploiting reduced order models, such as 1D blood flow simulations and LPMs[35–38]. Despite requiring less computational time, low dimensional models cannot capture the complex 3-dimensionalty of flow in coronary artery, which contributes to the determination of the pressure field. Grande et al.[39] have proposed a 1D-3D hybrid model, that solves the 3D NSE within the stenotic branch and uses a 1D reduced model for the rest of the coronary circulation. However, such hybrid model still requires up to 30 hours to run. Another strategy to limit computational cost while still reproducing the 3D features of coronary flow, is using steady BCs. Given that FFR is determined by the time-averaged ratio of Pa and Pd over the cardiac cycle, different authors have proposed to compute FFR-CT, relying on the solution of the 3D NSE imposing steady BCs[40–44]. Lo et al.[42] specifically compared the FFR-CT predicted by a transient and a steady simulation, obtaining a very good matching (average difference 1.5%), however their analysis was limited to a small cohort of patients (*n*=4). Liu et al.[41] observed a slightly higher diagnostic accuracy of FFR-CT predicted by the steady model vs. pulsatile, on a larger cohort (*n*=136) of patient with moderate stenosis. Their method yet yielded a moderate correlation in validation.
A robust, straightforward, and fully automated method for estimating FFR-CT from CCTA raw data, accounting for the physics of the problem, has not yet been described in the literature. In this work we present an automated time-efficient framework to perform CFD simulations of coronary flow from 3D CCTA, that introduces the following main improvements: *i)* a self-adjusting algorithm for setting patient-specific BCs; *ii)* an extensive assessment of FFR-CT obtained with steady-state vs. pulsatile CFD simulations on a large cohort of CAD patients;; *iii)* a validation of FFR-CT against invasive FFR; *iv)* a systematic characterization of the stenosis along the coronary tree, based on hemodynamics features that contribute to determining the risk of adverse events.

## 2) Methods

In the present work, we propose an automated and time-efficient framework for simulating coronary flow under different BCs. The entire adopted pipeline is schematized in **Figure 1**.

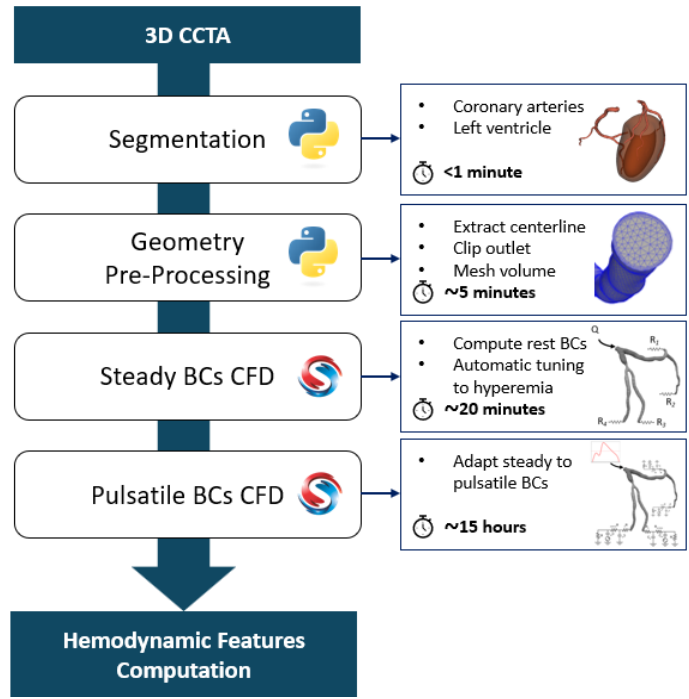

**Figure 1.** Schematic representation of the whole pipeline adopted in this study. Starting from volumetric CCTA images, the coronary lumen is automatically segmented; then the geometry is clipped with planes perpendicular to the centerlines, to generate inlet and outlet surfaces for CFD and meshed. Finally steady and pulsatile BCs are performed, and hemodynamic indexes are computed. Each box on the right indicates the time demand of the corresponding step in the workflow (referred to the machine used in this study).

### 2.1) Dataset

95 patients scheduled for clinically indicated invasive coronary angiography for suspected CAD were studied by CCTA by expert readers according to European Association of Cardiovascular Imaging (EACVI) guidelines[45,46] at Centro Cardiologico Monzino (Milan, Italy). Images were acquired with a GE Revolution CT machine (GE Healthcare, Milwaukee, Wisconsin), image dimension was 512×512×256 pixels, with pixel spacing ranging from 0.365×0.365 to 0.4×0.4 mm$^2$, and slice thickness of 0.4 to 0.65 mm. For each patient, invasive FFR was measured in at least one coronary branch. 133 mild-to-severe stenotic vessels were reconstructed from CCTA using an in-house automatic code based on a pre-trained 2-stage convolutional neural network[19] and thereafter manually refined. Segmentation included left and right coronary arteries downstream of the ostia, while excluding the aortic root. In addition, the left ventricular myocardial wall, which is later used for BCs setting, was automatically segmented with an open-source code[47]. An exemplifying geometry is showed in **Figure 2**. The clinical and demographics characteristics of the patient cohort are summed up in **Table I**. The present study was performed in accordance with recommendations of the local Ethics Committee (available in the **Supplementary Material**), with written informed consent from all subjects, in accordance with the Declaration of Helsinki.

**Table I**. Cohort characteristics.

| Characteristics | Data |
|---|---|
| Number of patients (*n*) | 95 |
| Number of vessels | 133 |
| Male/Female (*n*) | 71/24 (75%/25%) |
| Age (years) | 64.9 ± 8.4 |
| BMI (kg/m$^2$) | 26.7 ± 4.8 |
| Smoking (*n*) | 58 (44%) |
| Hypertension (*n*) | 86 (65%) |
| Diabetic (*n*) | 29 (22%) |
| Systolic Pressure (mmHg) | 141.0 ± 14.5 |
| Diastolic Pressure (mmHg) | 78.8 ± 8.1 |
| Heart rate (bpm) | 68.4 ± 9.8 |
| Stenosis grading (*n*) | |
| *Mild* | 91 (68%) |
| *Moderate* | 36 (27%) |
| *Severe* | 6 (5%) |

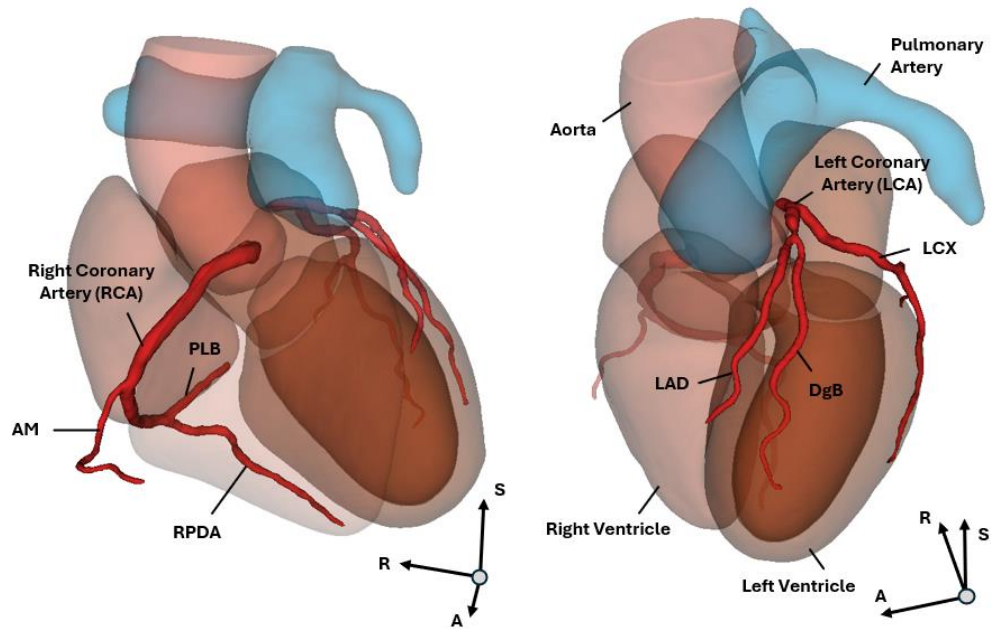

**Figure 2.** Example of reconstructed geometry of coronary arteries. Right coronary artery included the acute marginal (AM), the posterior descending artery (RPDA) and the postero-lateral (PLB) branches. Left coronary artery (LCA) included the left main trunk, the left anterior descending (LAD) artery, the circumflex artery (LCX) and the diagonal (DgB) and marginal branches if caliber was greater than 1.5 mm. Surrounding organs are not included in the CFD domain and are represented only to provide a reference.

## 2.2) Geometry pre-processing and meshing

After segmentation, the surface was automatically smoothed and remeshed (see **Supplementary Material**) and coronary centerlines were automatically extracted as described in our previous works[19,48] using the vascular modeling toolkit (VMTK) library[49] (**Figure 3.a**). To automatically generate inlet and outlet surfaces for CFD simulation, the geometry was clipped in correspondence of centerline end points with planes defined by the direction of the local centerline's Frenet tangent vector (**Figure 3.b**). If the diameter (defined through the maximum inscribed sphere) of the vessel at the end point was <1.5 mm[24], the centerline point where the clipping was performed was recursively shifted upstream along the vessel. The obtained geometry of the lumen was finally discretized into tetrahedral elements using SimVascular[50] embedded meshing algorithm (i.e., TetGen): global element size was set to 0.25 mm, with radius-based refinement in narrow regions and a boundary layer consisting of 3 prismatic elements was generated (**Figure 3.c**). The mesh element size was defined after a mesh sensitivity analysis (available in **Supplementary Material**).

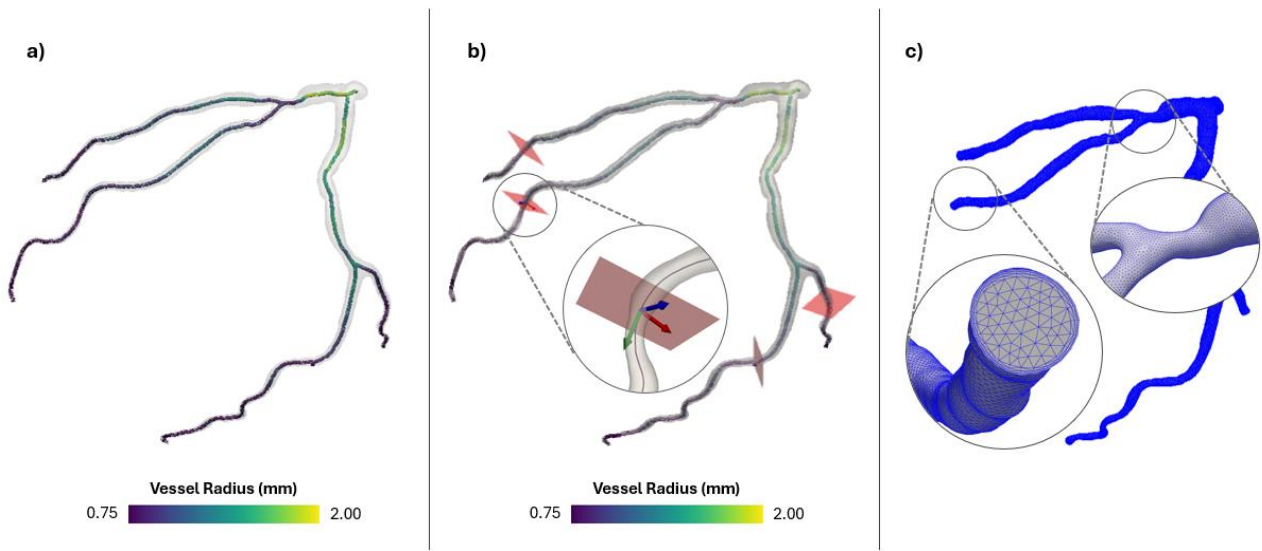

**Figure 3.** *a)* Example of centerline automatically extracted in left coronary artery; *b)* Detection of points where performing the clip of the surface to generate CFD outlets. The clipping was performed with planes (in red) defined by the Frenet tangent of the centerline (green vector); *c)* Mesh of the CFD model, showing details of a stenosis and the boundary layer generated.

## 2.3) Computational model of blood flow

Blood was assumed to be a homogeneous Newtonian fluid, with dynamic viscosity $0.004 \, \text{Pa} \cdot \text{s}$ and density $1060 \, \text{kg/m}^3$. Flow was assumed to be incompressible and laminar[42]. The vessel wall was assumed to be rigid, and a no-slip condition was applied. The governing equations (i.e., the 3D incompressible NSE) were numerically solved using the open-source software SimVascular[50]:

$$\nabla \cdot \mathbf{u} = 0$$

$$\rho \left( \frac{\partial \mathbf{u}}{\partial t} + \mathbf{u} \nabla \cdot (\nabla \mathbf{u}) \right) = -\nabla p + \mu \Delta \mathbf{u}$$

where **u** denotes the velocity field, ρ the fluid density, p the pressure and μ the dinamic viscosity. Body forces were neglected. Simulations were performed on a 40 cores Intel® Xeon® CPU X5670 machine with 64.4 GB RAM.

**2.4) Automatic tuning of steady BCs**

In our simulations with steady BCs, we adopted a constant flow rate at the inlet and resistance elements at the outlets. During hyperemia, coronary flow generally increases by 4-folds, while distal resistances decrease. On the other hand, the impact of vasodilators on blood pressure and heart rate can be neglected for the purpose of a CFD simulation[51]. Both flow and resistance values were computed starting from rest condition and then adjusted to account for the hyperemic state of the patient during invasive FFR evaluation.

First, mean aortic pressure ($P_{ao}$) was estimated from patient's systolic and diastolic brachial pressure (SP and DP) and heart rate (HR)[51]:

$$P_{ao} = DP + \left(\frac{1}{3} + 0.0012 \times HR\right) \times (SP - DP) \;[mmHg]$$

Coronary flow depends on the myocardial oxygen consumption and is primarily determined by patient's heart rate and pressure. We estimated the resting flow rate using the correction to Sharma et al.[51] framework proposed by Muller et al.[52]:

$$Q_{rest} = 0.14 \times (7 \cdot 10^{-4} \times HR \times SP - 0.4) \times M_{myo} \;[ml/min]$$

The myocardial mass was computed by adjusting the left ventricle mass with a gender-specific coefficient $M_{myo} = c \times M_{lv}$, with $c = 2.39$ and $2.34$ for men and women, respectively. The left ventricle mass was obtained through the left ventricle wall volume calculated from the CCTA segmentation, assuming a tissue density of 1.05 g/ml [53,54]. The total resting resistance was finally computed as $R_{rest} = P_{ao}/Q_{rest}$.

Then, the resting resistance was switched to hyperemic resistance $R_{hyp}$ by introducing the total coronary resistance index (TCRI)[51,55], that is defined as the ratio $R_{hyp}/R_{rest}$ and can be obtained as a function of resting HR:

$$TCRI = \begin{cases} 0.0016 \times HR + 0.1, & HR \leq 100 \\ 0.001 \times HR + 0.16, & HR > 100 \end{cases}$$

The hyperemic total resistance was then obtained as $R_{hyp} = R_{rest} \cdot TCRI$, and it was distributed to each outlet of the model on the basis of Murray's law[56], by which the flow rate through the *i*-th vessel is proportional to its radius to an exponent, that for coronary arteries is equal to 2.6, based on literature studies[24,57]: $Q_i \sim r_i^{2.6}$. Such relation leads to the follow, from which terminal resistances were determined:

$$Q_i = Q_{rest} \cdot \frac{r_i^{2.6}}{\sum_{k=1}^n r_k^{2.6}} \rightarrow R_i = \frac{P_{ao}}{Q_i} = P_{ao} \cdot \frac{\sum_{k=1}^n r_k^{2.6}}{Q_{rest} \cdot r_i^{2.6}}$$

The hyperemic flow rate was iteratively adjusted starting from a initial value, assumed to be $Q_{hyp,0} = 3.5 \cdot Q_{rest}$ in order to minimize the error on the resulting aortic pressure: briefly, during the run of the simulation, the relative difference $\epsilon_{Pao} = (P_{ao,CFD} - P_{ao})/P_{ao,CFD}$ and the inflow was consecutively adjusted as $Q_{hyp,n+1} = Q_{hyp,n} \cdot (1 + \epsilon_{Pao})$ (**Figure 4**), until a relative difference

$\epsilon_{Pao} < 2\%$ was obtained. The performance of the presented automated tuning pipeline for the flow rate was assessed comparing the FFR predicted using our method and Muller's expression for the inflow[52] in a subset of 20 patients.

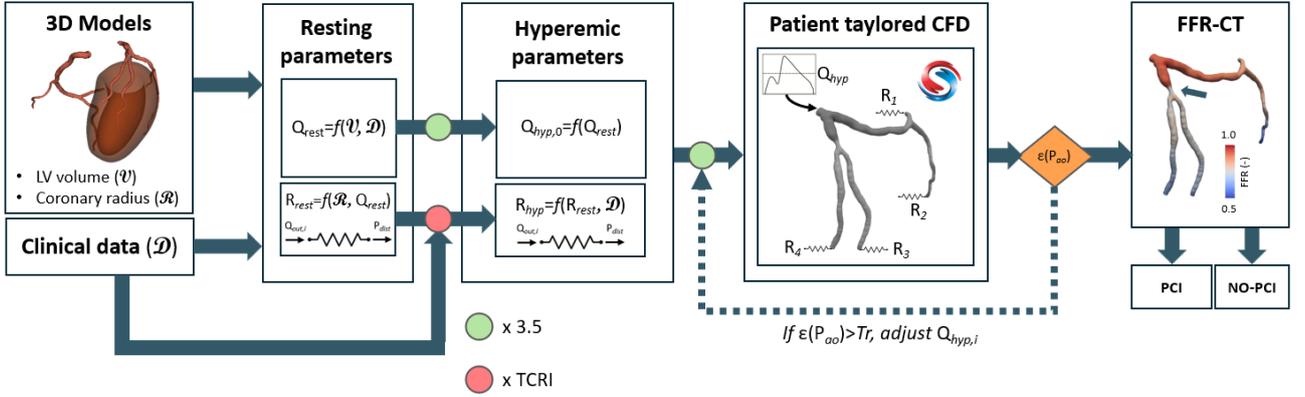

**Figure 4.** Schematic representation of the workflow adopted for the self-tuning CFD model. The tolerance threshold *Tr* adopted for the error on the mean aortic pressure predicted was 2%.

**2.5) Pulsatile BCs CFD simulations**

To define the pulsatile BCs for CFD simulations, the tuned flow rate value adopted for the steady BCs CFD was used. At the inlet, a flow waveform was adapted from literature[58], in order to match patient-specific HR and $Q_{hyp}$. Specifically, the waveform period was modified as follows:

$$T_{ps} = \overline{T} \cdot \frac{\overline{HR}[bpm]}{HR_{ps}[bpm]}$$

Where $T_{ps}$ and $HR_{ps}$ are the patient-specific period and heart rate, and $\overline{T}$ and $\overline{HR}$ are the population averaged period (1 s) and hear rate (60 bpm). While flow rate was modified in order to satisfy the following relation:

$$\frac{1}{T_{ps}} \int_0^{T_{ps}} Q_{TR}(t) dt = Q_{SS}$$

Where $Q_{TR}(t)$ is the pulsatile flow waveform and $Q_{SS}$ is the flow rate value to which the steady CFD converged. Finally, at each outlet of the model the typical coronary 5-elements Winkessel model (5WK) was coupled (**Figure 5**). The tuning of the resistances and compliances was achieved using the framework described by Sankaran et al.[57]. The intramyocardial compliance using a ventricular pressure waveform scaled on patient's heart rate and systolic pressure.

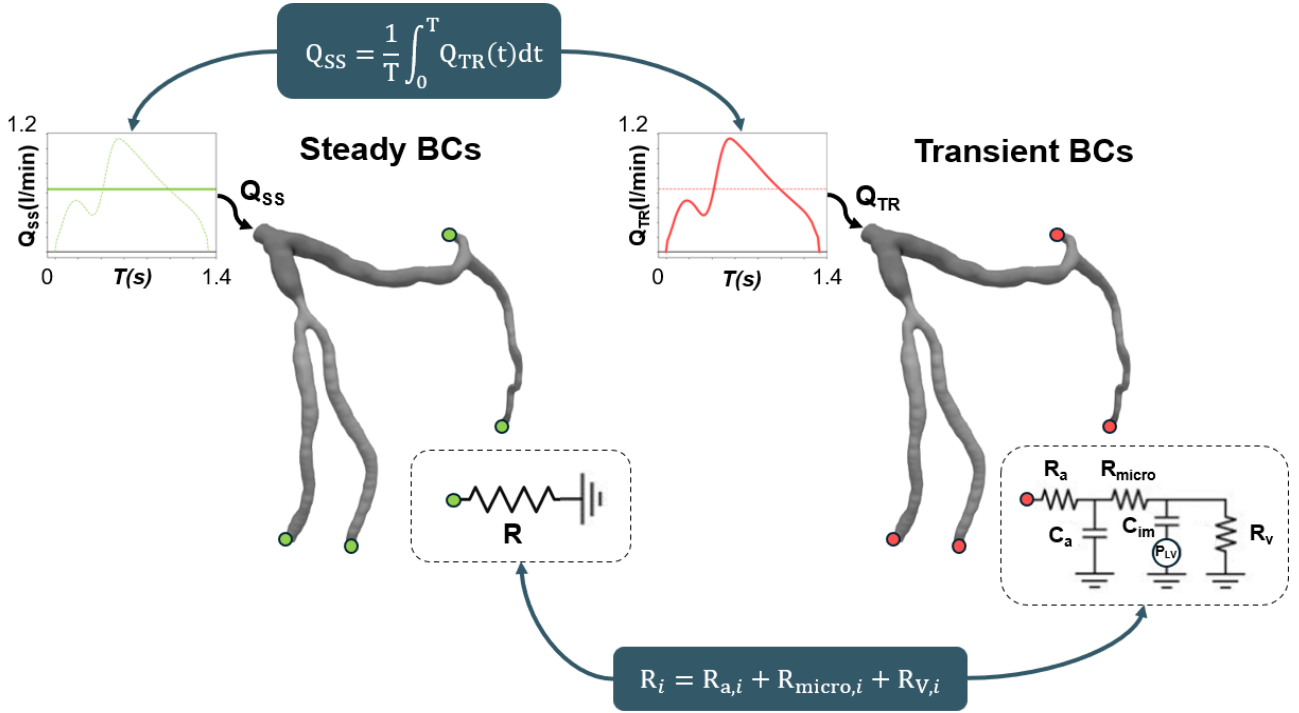

**Figure 5.** Representation of steady and transient boundary conditions adopted in the two modeling approaches and their relationship. Transient inflow is such that its time integral is equal to the steady inflow. At each branch, the resistances in the 5-elements Windkessel model are such that the sum is equal to the resistance in steady BCs model. Compliance and left ventricular pressure are defined for the transient case only.

**2.6) Calculation of hemodynamic indexes**

In steady BC simulations, once convergence was achieved, FFR-CT was point-wisely computed in the whole domain by dividing the local pressure by the pressure obtained at coronary ostium. In pulsatile flow simulations, pointwise FFR-CT was obtained computed as:

$$\text{FFRCT}_{TR}(\mathbf{x}) = \frac{1}{T}\int_0^T \frac{P(\mathbf{x},t)}{P_{ao}(t)} dt$$

Where $P(\mathbf{x},t)$ is the pressure at the generic location $\mathbf{x}$, $P_{ao}(t)$ the pressure at the ostium and $T$ the length of the cardiac cycle. The wall shear stress (WSS) was computed on the outer surface of each simulated model according to Newton's definition:

$$\text{WSS} = \mu \nabla_\mathbf{n} \mathbf{u}$$

Where $\mathbf{n}$ is the normal direction to surface. Additionally, the FFR gradient per millimeter (FFR-grad), along the vessel centerline, was computed to identify the region of functional CAD, that corresponds to regions with FFR drop $\geq 0.0015/\text{mm}$[59]. FFR-grad was obtained by projecting the pressure field onto the coronary centerline (based on minimum distance criterion), that was resampled to have a homogeneous spacing between points equal to 1 mm, and then computed as difference between the FFR value at each point and the subsequent and serves to determine the length of functional CAD[60]. Finally, for each coronary branch the pullback pressure gradient index (PPG)[59–62] was computed. PPG is an index between 0 and 1, that permits to classify the CAD pattern: a PPG close to 0 indicates diffuse CAD, whereas a PPG close to 1 focal CAD. It is defined as follows:

$$\text{PPG} = \frac{1}{2} \times \left\{ \frac{\text{MaxPPG}_{20mm}}{\Delta\text{FFR}_{vessel}} + \left(1 - \frac{\text{lenght of functional disease [mm]}}{\text{length of the vessel [mm]}}\right) \right\}$$

At each centerline point, the maximal PPG (MaxPPG$_{20mm}$) is defined as the maximum FFR drop over the adjacent 20 mm tract along the vessel, the ΔFFR of the vessel is the difference between the FFR at the ostium and at the most distal vessel anatomical location.

## 2.7) Detection and characterization of lesions

To detect and characterize possible culprit lesions, the centerlines of the main coronary vessels (i.e., the left anterior descending, the left circumflex and the right coronary artery) were isolated. Candidate lesions were identified by retrieving the points of local minimum in the centerline abscissa-vessel radius graph (**Figure 6**). The Python SciPy 1.9.1 signal library[63] was used for the analysis, setting a minimum value of width and prominence equal to 2 mm and 0.65 mm (corresponding to 3 and 1 pixel, for images with lower resolution, respectively). Lesions were characterized in terms of extent and prominence. Additionally, the maximum WSS along the lesioned tract and the trans-lesional $\Delta\text{FFR}_{les}$ (i.e., the difference between the FFR-CT at the end and at the beginning of the lesion) were computed.

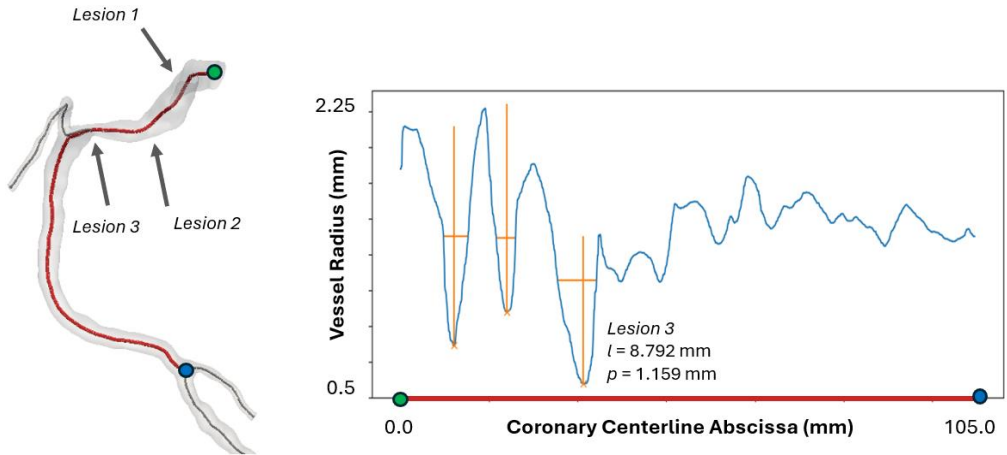

**Figure 6**. Example of right coronary artery isolated for processing, highlighted in red (left) and resulting centerline abscissa-vessel radius plot (right), where 3 candidate culprit lesions are identified by "x" mark, extent (*l*) and prominence (*p*) of the lesion are represented by horizontal and vertical segments, respectively, in the plot. The characteristic of one detected lesion are indicated, as an example.

## 2.8) Statistical analysis

Statistical analysis was performed using the Python SciPy 1.9.1 statistics library[63]. The relationship between invasive FFR (used as gold standard) and FFR-CT, computed both with steady and pulsatile BC CFD simulations, was examined by performing a linear regression analysis. The agreement between the methods was assessed by means of Bland-Altman plots. Diagnostic accuracy of the two numerical methods was evaluated though confusion matrix analysis in terms of precision, recall and accuracy and comparing the receiver operating characteristic (ROC) area under the curve (AUC).

Data normality was assessed using the D'Agostino-Pearson test, normally distributed data are reported as mean (±standard deviation), non-normally distributed data are reported as median (interquartile range). For all the tests, a *p* value below 0.05 was considered statistically significant.

## 3) Results

### 3.1. Computational performances

The time performance of the deployed automatic framework for coronary CFD was defined as the time to obtain results, starting from 3D CCTA images. On average, the time required for the segmentation and the preprocessing (consisting in centerline extraction, outlet preparation and volume meshing) of the vessel geometry was <10 minutes. Time-gain was quantified by comparing the duration of the same process performed by three independent expert operators (i.e., three bioengineers with 7, 3 and 2 years of experience). The manual preprocessing took on average 2-to-4 hours, thus the time-gain is quantifiable as a reduction of approximatively 93% of the preprocessing time. Steady BC CFD simulations took between 15 and 40 minutes, while pulsatile BC CFD simulations calculation time was 8 to 14 hours, thus the cost reduction was approximately 96%. Thus, the overall computational cost reduction achieved was of 95% on average.

In steady BC simulations, the automatic tuning of hyperemic flow rate converged within 2 flow rate adjustments. The average flow rates for right and left coronary artery were 626 (±194) ml/min and 460 (±162) ml/min, respectively, which align to realistic hyperemic flow rate population-average values (626.32 and 453.68 ml/min for left and right coronary, respectively)[52].

### 3.2 Comparison of FFR, FFR-CT steady and FFR-CT transient

To compare the results of the steady and transient BC simulations, a time-average filter was applied to the solution of the transient BC CFD. **Figure 7** reports the pressure and FFR-CT field resulting from the steady and transient BC simulations (FFR-CT$_{SS}$ and FFR-CT$_{TR}$), as well as a map of the pointwise relative difference between the two, for 6 cases. The average pointwise pressure relative difference obtained was -0.19% (±8.4%), while the average of the pointwise FFR-CT$_{SS}$ and FFR-CT$_{TR}$ relative difference (i.e., (FFR-CT$_{SS}$ – FFR-CT$_{TR}$)/FFR-CT$_{TR}$) was -1.73% (±1.1%). Focusing on the lesion, the median value (interquartile range) of the invasive FFR, FFR-CT$_{SS}$ and FFR-CT$_{TR}$ were 0.83 (0.77-0.90), 0.84 (0.79-0.91), 0.83 (0.77-0.89), respectively. **Figure 8a** shows a scatter plot of invasive FFR and FFR-CT$_{SS}$. A moderate-to-strong positive correlation resulted from linear regression analysis with $r$=0.7972 ($R^2$=0.634, $p$<0.001). The Bland-Altman plot for FFR and FFR-CT$_{SS}$ is reported in **Figure 8d**. On average, FFR-CT$_{SS}$ exceeded invasive FFR by 0.0130 (95% limit of agreement -0.1502 to 0.1282), indicating a good agreement between the two measurements. Similarly, FFR-CT$_{SS}$ was compared versus FFR-CT$_{TR}$: linear regression (**Figure 8b**) indicated a strong positive correlation, with $r$=0.9881 ($R^2$=0.976, $p$<0.001), while from the Bland-Altman analysis (**Figure 8e**) the FFR-CT$_{SS}$ exceeded FFR-CT$_{TR}$ by 0.0118 (95% limit of agreement -0.0146 to 0.0383). Finally, FFR and FFR-CT$_{TR}$ were compared. A slightly weaker correlation was obtained from linear regression (**Figure 8c**), with $r$=0.7612 ($R^2$=0.579, $p$<0.001) compared to FFR versus FFR-CT$_{SS}$ comparison. The Bland-Altman analysis (**Figure 8c**) produced a lower bias (i.e., -0.0045) with a slightly larger confidence interval (0.253 compared to 0.243). The relative error, between FFR-CT$_{SS}$ and FFR was 2.09%, while the error FFR-CT$_{TR}$ vs. FFR-CT$_{SS}$ was 1.76%.

In the subset of 20 patients used to compare our methods for defining BCs to literature-based method, the average error between FFR-CT$_{SS}$ and FFR was 4% when using our method and 6% when using the relation from the literature, however, mean aortic pressure settled to patient specific values only when using our method for determining BCs (median (interquartile) of pressure achieved with clinical measure, our approach and Muller's were 97.5 (86, 113.5), 98.5 (87, 115.3) and 134 (108.6, 241.2) mmHg, respectively).

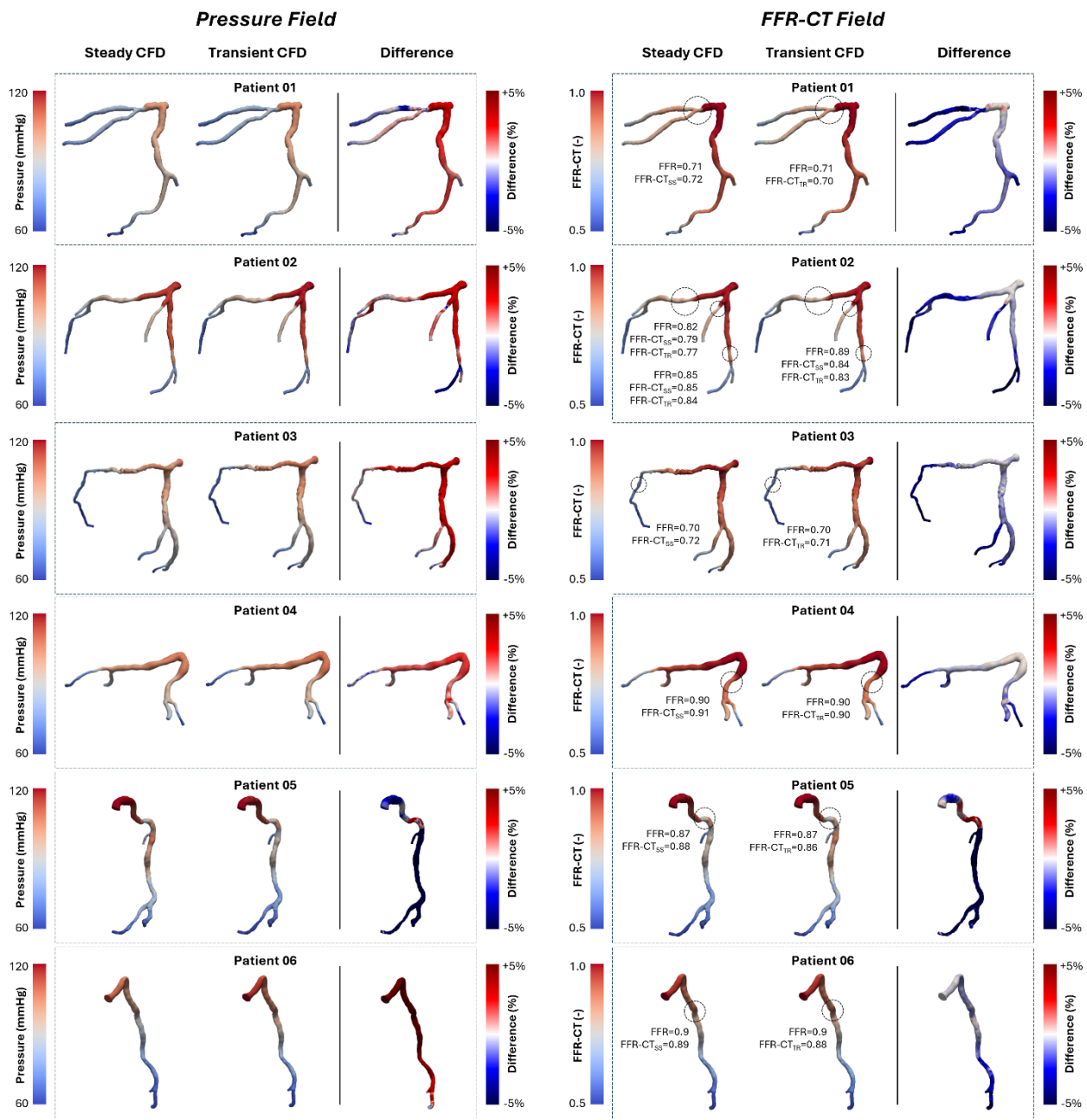

**Figure 7**. Contour plots of pressure and FFR-CT field for 4 representative left branches (Patient 01-04) and 2 right branches (Patient 05-06). For each patient, from left to right are the field predicted with steady CFD, the field predicted with transient CFD and the per-point relative difference between the two scalar fields (i.e., (scalar$_{SS}$ – scalar$_{TR}$)/scalar$_{TR}$). Patient 4 is the case for which best agreement between FFR-CT$_{SS}$ and FFR-CT$_{TR}$ was achieved (average per-point difference was -0.22%), Patient 5 is the case of worst agreement (average per-point difference was -5.80%).

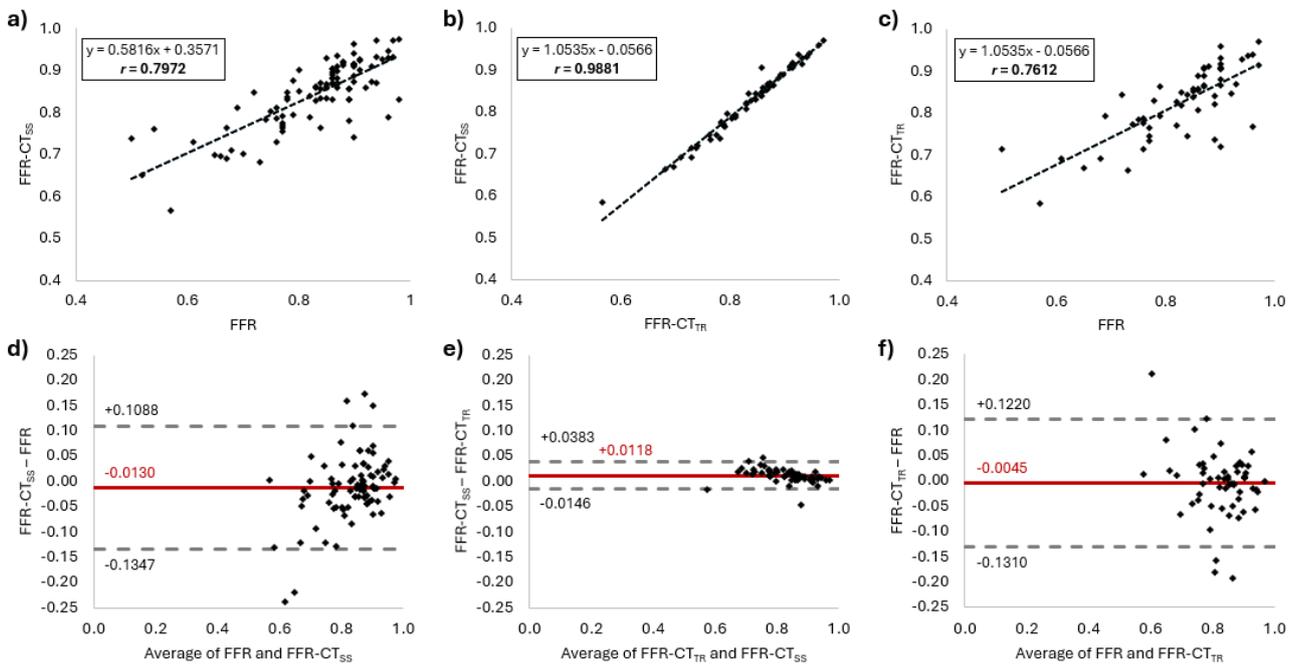

**Figure 8**. Scatter plot an Bland-Altman analysis illustrating the correlation between FFR and FFR-CT steady.

### 3.3) Diagnostic Accuracy

To evaluate the diagnostic performance of the implemented numerical framework, the invasive measure of FFR was adopted as diagnostic criterion. Precision and recall for FFR-CT$_{SS}$<0.8 versus FFR<0.8 were 0.808 and 0.701, respectively, with an overall diagnostic accuracy of 0.862. Precision, recall and overall accuracy for FFR-CT$_{TR}$<0.8 versus FFR<0.8 were 0.773, 0.850 and 0.864, respectively. The areas under the curve (AUC) in the ROC receiver characteristic curve for FFR-CT$_{SS}$ and FFR-CT$_{TR}$ were 0.923 and 0.912, respectively (**Figure 9**), thus a very good diagnostic performance was achieved in both cases. The Youden's J index was 0.736 and 0.697 for FFR-CT$_{SS}$ and FFR-CT$_{TR}$, respectively, corresponding to a cut-off value of 0.81 and 0.80.

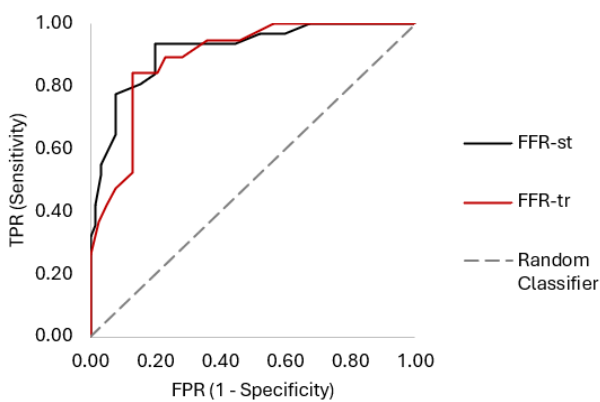

**Figure 9**. Receiver characteristic curve (ROC) for FFR-CT computed with steady (st) and transient (tr) CFD simulations. Invasive FFR<0.8 was adopted as diagnostic criteria.

**Table II.** The table sums up the performance metrics achieved with the two methods (bold font indicates the highest value). The $\Delta_{SS-TR}$ denotes the relative difference (i.e., (metric$_{SS}$ – metric$_{TR}$)/metric$_{TR}$) achieved for each specific performance.

|  | FFR-CT$_{SS}$ | FFR-CT$_{TR}$ | $\Delta_{SS-TR}$ |
|---|---|---|---|
| Accuracy | 0.860 | **0.864** | -0.46% |
| Precision | **0.808** | 0.739 | +9.34% |
| Recall | 0.700 | **0.850** | -17.6% |
| NVP | 0.878 | **0.930** | -5.59% |
| Specificity | **0.929** | 0.870 | +6.78% |
| AUC | **0.923** | 0.912 | +1.21% |

### 3.4) Hemodynamic characterization of lesions

For each of the main coronary branches (i.e., the left anterior descending, the left circumflex and the right coronary artery) suspect lesions were identified as described in **Section 2.7**. To determine the diagnostic predictability of different hemodynamic descriptors associated with the lesion, these were compared for cases in which PCI was recommended and not, based on the invasive FFR value (i.e., if the FFR value was below 0.80 or not). The maximum WSS in the lesion area was significantly higher ($p=0.0365$) for the PCI group (**Figure 10a**), while no relevant difference was obtained for average WSS. Trans-lesional FFR, resulted significantly higher ($p=0.0010$) in the PCI group (**Figure 10b**). No significant difference was obtained between lesion length and prominence in the PCI versus no-PCI group (**Figure 10c-d**). The correlation between each hemodynamic feature and FFR was assessed through Pearson correlation analysis. As reported in **Figure 10f**, none of the variables exhibited a moderate or strong correlation with FFR. As proposed by Collet et al.[60], each lesion was classified based on the PPG tertiles T1 and T2: lesions with PPG<T1 were considered diffuse, while lesions with PPG>T2 were considered focal. The cut-off T1 and T2 values were 0.61 and 0.74, which are slightly higher than those obtained by Collet et al. in their study (0.55 and 0.71), on a dataset of 158 vessels. On average, lesions with low and intermediate values of PPG were associated with lower values of FFR despite no statistically relevant difference was obtained ($p>0.05$). Comparing the computed PPG values for lesions which required PCI versus lesions that were not treated (**Figure 10e**), PPG resulted significantly lower in the PCI group ($p=0.0101$). **Table III** sums up the results achieved from PPG characterization.

**Table III**. Clinical baseline and procedural characteristics stratified by pullback pressure gradient (PPG). *n* denotes the number of cases.

| Characteristics | Low PPG (Diffuse lesion) | Intermediate PPG | High PPG (Focal lesion) |
| --- | --- | --- | --- |
| PCI (*n*) | 28 (47.8%) | 15 (30.0%) | 3 (10.0%) |
| No PCI (*n*) | 30 (52.2%) | 35 (70.0%) | 23 (90.0%) |
| FFR (-) | 0.81 (±0.10) | 0.82 (±0.09) | 0.88 (±0.10) |

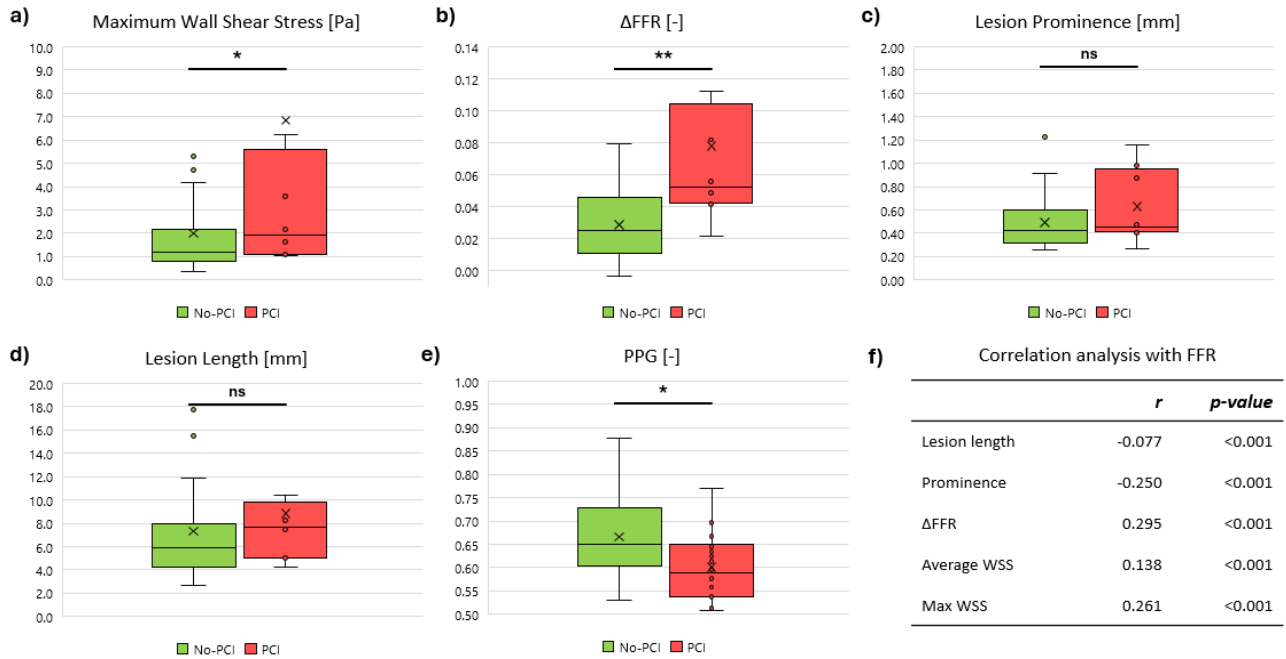

**Figure 10.** Comparison of different hemodynamic features in lesions for which percutaneous intervention (PCI) was performed and not. The "ns", "*", "**" denotes a p-value >0.05, <0.05 and <0.01, respectively.

## 4) Discussion

FFR is an established index for CAD assessment. Along with other functional features of lesions, such as PPG, trans-lesional FFR and geometric parameters, it provides valuable information to cardiologist, for effective diagnosis. CFD models enables non-invasive computation of FFR-CT and other hemodynamic indexes, however the deployment of a patient-tailored model of coronary flow requires the tuning of BCs that rarely are measured *in-vivo* and entails a significant computational cost due to time-varying flowrate. In the proposed study, a time-efficient framework for the automatic computation of FFR-CT and the hemodynamic characterization of coronary lesions was developed and tested on a dataset of 133 coronary vessels reconstructed from 95 patients referred to CCT for suspected CAD. The developed methodology relies on steady BC CFD simulation of coronary flow and exploits a fully automated pipeline that does not require specific expertise or operator inputs. The key potential of the described framework lies in its fully automated methodology, which prevents the introduction of possible biases in the analysis. This approach provides an accurate estimation of the FFR in a short time without requiring invasive measurement, ensuring a good trade-off between computational cost and diagnostic accuracy. In summary, the main innovative aspects of the present study are: i) the deployment of a fully automated and computationally efficient framework for computing FFR-CT, validated against invasive measurements on a large cohort of patients; ii) the systematic comparison of steady and transient BC CFD simulations for FFR-CT estimation; iii) a validation for FFR-CT computed with, less time demanding, steady BCs CFD simulations; iv) the comparison of hemodynamic features in coronary lesion that required PCI versus lesions that did not.

### 4.1) Automatic framework performance

The benefits of using an automated framework for the preparation of the CFD models are evident in terms of time saved. The use of a deep learning-based segmentation algorithm, that once trained, systematically produces the same result for a given input, also prevents from introducing inter- and intra-operator bias. The computational demand of steady BC simulations compared to pulsatile ones, is significantly lower; additionally, using a steady framework, does not require the tedious process of tuning a complex LPM (such as the 5WK) for outlet boundary conditions, as it uses simple resistances. Some authors have proposed more time-efficient solutions[38,64], fully based on deep learning models, which however do not consider for the complete 3-dimensional coronary artery hemodynamics. In our framework, the complete coronary vasculature is analyzed, nonetheless providing a solution in a relatively short time.

In general, tuning BCs is a nontrivial process, Muller et al.[52] conducted a thorough analysis on the impact of different BCs definition strategies described in the literature: our method exploits patient-specific information for the definition of distal resistances and a self-adjusting algorithm that allows for automatic tuning of inflow rate in other to match patient's pressure at the inlet. When comparing the error obtained in a subset of 20 patients for inflow BC setting using our method and Muller's equation[52], the error with respect to invasive measure of FFR was 4% and 6%, respectively. Despite the comparable error achieved, the pressure obtained at the coronary ostium matched the patient pressure only when using our approach and reached non-physiological values when using literature formula.

## 4.2) Comparison of steady-CFD and transient-CFD for FFR estimation

**Table IV** to **VI** sum up the comparison of the results achieved in our work with previously published studies in terms of steady vs. transient BCs comparison, validation vs. FFR and diagnostic performance of the methodology.

**Table IV.** Comparison between FFR-CT$_{SS}$ and FFR-CT$_{TR}$. $n$ = number of cases. Pearson $r$ = Pearson correlation coefficient (FFR-CT$_{SS}$ vs. FFR-CT$_{TR}$). Bias = FFR-CT$_{SS}$ – FFR-CT$_{TR}$. LoA = Limit of Agreement. Relative difference = (FFR-CT$_{SS}$ – FFR-CT$_{TR}$)/FFR-CT$_{TR}$. Bold font indicates the best metric value obtained.

| FFR-CT$_{SS}$ vs. FFR-CT$_{TR}$ | $n$ | Pearson $r$ | Bias | 95% LoA | Relative difference |
|---|---|---|---|---|---|
| Alzhanov et al.[40] | 3 | - | - | - | -1.9% |
| Liu et al.[41] | **136** | 0.75 | **0.01** | [-0.17; 0.20] | 6.7% |
| Lo et al.[42] | 4 | - | - | - | 1.3% |
| Shi et al.[65] | 29 | - | - | - | **-0.1%** |
| **Our method** | 133 | **0.988** | **0.01** | **[-0.02; 0.04]** | 1.76% |

**Table V.** Comparison between FFR-CT and FFR. $n$ = number of cases. $r$ = Pearson correlation coefficient (FFR-CT$_{SS}$ vs. FFR-CT$_{TR}$). Bias = FFR-CT$_{SS}$ – FFR-CT$_{TR}$. LoA = Limit of Agreement. Error = (FFR-CT – FFR)/FFR. Bold font indicates the best metric value obtained.

| | | *Steady BCs CFD* | | | | *Transient BCs CFD* | | | |
|---|---|---|---|---|---|---|---|---|---|
| **FFR-CT vs. FFR** | $N$ | $r$ | Bias | 95% LoA | Error | $r$ | Bias | 95% LoA | Error |
| Alzhanov et al.[40] | 3 | - | - | - | **0.2%** | - | - | - | **-2.2%** |
| Liu et al.[41] | **136** | 0.75 | -0.03 | [-0.24; 0.18] | 11% | - | - | - | - |
| Lo et al.[42] | 4 | - | - | - | -14% | - | - | - | 15.1% |
| Shi et al.[65] | 29 | 0.729 | 0.03 | [-0.08; 0.14] | -3.8% | 0.742 | 0.03 | **[-0.07; 0.14]** | -3.7% |
| **Our method** | 133 | 0.797 | **0.01** | [-0.13; 0.10] | 2.1% | **0.761** | -0.005 | [-0.13; 0.12] | 2.2% |

    **Steady vs. transient.** The very strong correlation ($R^2$=0.976, $p$<0.001) achieved between FFR-CT values predicted with steady and transient BC CFD simulations indicates that FFR-CT can be computed exploiting time-averaged boundary conditions and less computationally demanding steady simulations. Despite being already addressed by several authors[40–42,65,67], the present study reports, to the best of our knowledge, the highest correlation coefficient achieved between the two methods in a population study. The Bland-Altman analysis yielded a bias between FFR-CT$_{SS}$ and FFR-CT$_{TR}$ of 0.0118, in agreement with the results reported in the work of Liu et al.[41] The average error between FFR-CT$_{SS}$ and FFR-CT$_{TR}$ resulted 1.76%, which is comparable to values achieved by other authors (**Table IV**). The good agreement between the two methods, steady and transient BC CFD, was assessed in the whole computational domain: indeed, the pointwise difference of pressure and FFR-CT fields yielded average values below 1% and 2%, respectively (**Figure 7**).

**Table VI.** Comparison between diagnostic performance of different methods. $n$ = number of cases. Acc = Accuracy. Prec = Precision. Spec = Specificity. AUC = Area under the ROC curve. Bold font indicates the best metric value obtained.

| Steady BCs | n | Acc | Prec | Recall | Spec | AUC |
|---|---|---|---|---|---|---|
| Liu et al.[41] | **136** | 0.883 | **0.828** | 0.681 | 0.934 | 0.85 |
| Shi et al.[65] | 29 | 0.750 | 0.667 | 0.579 | **0.941** | 0.88 |
| **Our method** | 133 | **0.860** | 0.808 | **0.700** | 0.929 | **0.923** |
| Transient BCs | n | Acc | Prec | Recall | Spec | AUC |
| Coenen et al.[9] | 189 | 0.746 | 0.648 | 0.813 | 0.651 | 0.833 |
| Liu et al.[41] | 136 | **0.876** | **0.820** | 0.681 | **0.945** | 0.81 |
| Shi et al.[65] | 29 | 0.806 | 0.727 | 0.684 | 0.941 | **0.93** |
| **Our method** | 133 | 0.864 | 0.739 | **0.850** | 0.870 | 0.912 |

**Validation.** Both methods, FFR-CT$_{SS}$ and FFR-CT$_{TR}$, were validated against invasive FFR, yielding a strong correlation ($r_{SS}$=0.7972, $r_{TR}$=0.7612) with values of $r$ coefficient comparable to those obtained by other authors[9,41,65] (**Table V**). Interestingly the correlation between FFR-CT$_{SS}$ and FFR resulted stronger than FFR-CT$_{TR}$ and FFR. On the other hand, the Bland-Altman analysis yielded a lower bias comparing FFR-CT$_{TR}$ and FFR with respect to FFR-CT$_{SS}$ and FFR, -0.0045 vs. -0.0130, outperforming the results reported in previous works[9,21,41,65]. The confidence interval was slightly larger for FFR-CT$_{TR}$-FFR than FFR-CT$_{SS}$-FFR, with values of 0.253 and 0.243, respectively.

**Diagnostic accuracy.** The AUC was used to evaluate the diagnostic performance for each stenotic vessel. The FFR-CT$_{SS}$ exhibited a higher performance for the diagnosis of PCI compared to FFR-CT$_{TR}$, with a larger AUC, as reported by Liu et al.[41]. Overall, our results align with those reported in similar works (**Table VI**) in terms of accuracy, precision, recall and AUC [9,41,43,65]. The optimal cut-off value for FFR-CT$_{SS}$ was 0.81, which is slightly higher than the cut-off value indicated by consensus guidelines (i.e., 0.80), possibly due to fact that in steady BC CFD models, pressure loss due to flow acceleration are neglected and thus the pressure values tend to be slightly overestimated. The values of accuracy, precision and recall achieved are comparable to those reported in other studies[9,41,43,65].

## 4.3) Hemodynamic characterization of coronary lesions

Lesions detected within the coronary tree were characterized based on hemodynamic features. Cases that required PCI exhibited both a higher WSS and trans-lesional FFR, which is coherent with the results of the EMERALD[68] trials, that proved that both parameters carry major information in the identification of the culprit lesion and confirms that WSS and trans-lesional FFR could enhance the diagnostic process of CAD. Moreover, high WSS is an index associated with atherosclerosis progression and rupture risk[69–72], thereby not only can aid in the diagnosis CAD, but also in the ongoing monitoring of high-risk cases. Based on the PPG values, diffuse lesions required surgical intervention significantly more frequently than focal lesions. Despite focal lesions tended to produce a larger drop pressure, the downstream expansion produces a pressure recovery that is totally absent in the case of diffuse lesion along the whole vessel. A similar result is reported by Collet et al.[60], which observed a higher frequency of major adverse events and PCI in low PPG vessels. The morphological characteristics of the lesions showed no relevant differences between vessels necessitating PCI and not, consistently with the findings of several studies indicating that stenosis alone is not the strongest indicator for CAD diagnosing[8,10,11,65]. Furthermore, Wu et al.[73] examined the correlation between the degree of stenosis and FFR, finding a moderate negative correlation ($r=-0.328$, $p<0.001$). On the other hand, the stenosis grading and extent of the lesion concur in the assessing of clinical risk[68] and should not be disregarded. In our study, we investigated the relationship between the hemodynamic descriptors analyzed and FFR, which is a strong independent indicator of CAD. None of the descriptors exhibited a strong correlation with FFR, suggesting that while such indexes can be used to deepen the analysis of CAD, they don't serve as independent indicators of the pathology.

**4.4) Limitations and future work**

The present work is not exempt from limitations. First, the retrieved dataset is not balanced: the number of patients with severe stenosis is limited, and such patients are the ones on which it is paramount that precision of the method is maximum. A larger number of subjects with severe stenosis should be included in the dataset for a more comprehensive analysis. While the time gain is evident compared to manual segmentation, using deep learning models may introduce errors in the reconstruction of geometry, affecting the accuracy of the CFD results. In our study, we demonstrate that this approach still yields good diagnostic performance, as shown by the ROC analysis, and notably reduces overall computation time. The anatomical model was truncated to retain only vessels that are clinically relevant for FFR assessment (i.e., with caliber > 1.5 mm). Despite this representing a limitation for our method, as part of the coronary tree is not simulated, Shi et al.[26] showed that the time-averaged pressure within the coronary artery tree is not significantly affected by the removal of any sub-branch from the anatomical model, and consequently the FFR-CT estimation. Thus, removing small distal branches, that favors numerical convergence, does not introduce a significant bias in our framework. Since coronary flow measurement was not obtainable from any exam, our framework estimates the inlet flow rate by adapting it to the patient's pressure. This introduces some uncertainty in replicating the patient-specific inflow; nevertheless, the sensitivity of FFT-CT to small variations in flow rate is limited, as demonstrated by the good agreement achieved by our method with invasive FFR measurements. Recently, Xue et al.[66] proposed a more sophisticated approach to estimate coronary inflow, yielding promising results in terms of resulting FFR-CT. Finally, the time required for computing FFR-CT$_{SS}$, despite being significantly lower than using transient BC CFD, is still about 30 minutes. In cases in which such time may not be available, a more efficient solution must be adopted. In future studies, our aim will be to predict the coronary pressure field directly, exploiting a deep learning neural network trained on the CFD models produced in this work, thus limiting to few minutes the calculation time.

**5) Conclusion**

We presented a time efficient framework for accurate and non-invasive calculation of FFR-CT. The methodology was compared to a traditional pulsatile flow models and validated on a cohort of 133 cases for which invasive FFR was available. The proposed approach exhibited a good diagnostic performance (AUC=0.932) and proved to be equivalent to a pulsatile model, for the calculation of time-averaged parameters. The methodology is highly automated and time efficient (20 minutes approximatively), encouraging its clinical applicability for the diagnosis of CAD and guidance of PCI surgery.

**Acknowledgments**

This work has been supported by Fondazione Regionale per la Ricerca Biomedica (Regione Lombardia). Project 3432721 - AI-CORPS. The publication was made possible with funding from Ministero dell'Università e della Ricerca as part of the PNC. Project ANTHEM.